\documentclass[10pt,letterpaper,twocolumn]{article} 

\usepackage{ol2}
\usepackage[draft]{hyperref}
\usepackage{amsmath}

\begin{document}
\twocolumn[ 

\title{{Two-photon transitions driven by a combination of \\ diode and femtosecond lasers}}

\author{Marco~P.~Moreno${^1}$, Giovana~T.~Nogueira${^2}$, Daniel~Felinto${^1}$ and Sandra~S. Vianna${^{1,*}}$}

\affiliation{$^1$Departamento de F\'{\i}sica,
Universidade Federal de Pernambuco,
50670-901 Recife, PE - Brazil \\
$^2$Departamento de F\'{\i}sica, Universidade Federal de Juiz de Fora,
36036-330 Juiz de Fora, MG - Brasil \\
$^*$Corresponding author: vianna@ufpe.br}

\begin{abstract}
We report on the combined action of a cw diode laser and a train of ultrashort 
pulses when each of them drives one step of the 5S-5P-5D two-photon transition in rubidium vapor.
The fluorescence from the $6P_{3/2}$ state is detected for a fixed repetition rate of the 
femtosecond laser while the cw-laser frequency is scanned over the rubidium $D_{2}$ lines. 
This scheme allows for a velocity selective spectroscopy in a large spectral range including the 
$5D_{3/2}$ and $5D_{5/2}$ states. The results are well described in a simplified frequency 
domain picture, considering the interaction of each velocity group with the cw laser and a single mode of the frequency comb.

\end{abstract}


] 

\maketitle

In recent years, advance in ultrafast lasers and related optical technologies has enhanced
the ability to control the interaction between light and matter\cite{Ye2005}. In particular, the application of 
mode-locked lasers for high-resolution precision spectroscopy has been explored using
fully stabilized optical frequency combs\cite{Jones2000}, as a direct probe of atomic transitions~\cite{Stowe2008a}. 
By incorporating pulse shaping technology, a precise control of molecular dynamics has also been demonstrated\cite{Stowe2008}.

For experiments along these lines focused on two-photon transitions, a cooled and trapped sample is usually employed, 
such that only one group of atoms is investigated as the fs laser repetition 
rate is scanned\cite{Stowe2008,Marian2004}. For atomic systems with considerable Doppler broadening,  
excitations to different states have been isolated by applying narrowband interference filters in the exciting-lasers pathways
\cite{Stalnaker2010}. In this context, the introduction of a second, cw laser 
opens new directions of investigation, with the narrowband laser assuming the role of a velocity-selective filter. In the case of one-photon transitions, velocity-selective spectroscopy has already been performed\cite{Aumiler2005}, 
and distinction between different hyperfine levels within the Doppler profile has been demonstrated\cite{Polo2011}.

In this work, the scheme with a second, cw laser is applied to investigate two-photon transitions, achieving considerably improved spectral resolution when compared to the previously reported works on velocity-selective spectroscopy with frequency combs~\cite{Aumiler2005,Polo2011}. We employ a diode laser and a 1 GHz Ti:sapphire laser to excite each step
of the 5S-5P-5D two-photon transition in a rubidium vapor (Fig. 1). The fluorescence of the $6P_{3/2}$
state is detected for a fixed repetition rate of the femtosecond (fs) laser as a function of the diode frequency.
This scheme allows us to performe a velocity-selective spectroscopy and investigate, simultaneously, the 
hyperfine levels $5D_{3/2}$ and $5D_{5/2}$. Under 
the high repetition rate of the fs laser, the necessary condition 
for the accumulation of population and coherence is easily fulfilled, and a good description of 
the results is obtained considering a three-level cascade system interacting with two cw lasers.

\begin{figure}[htbp]
  \centering
  \includegraphics[width=6.5cm]{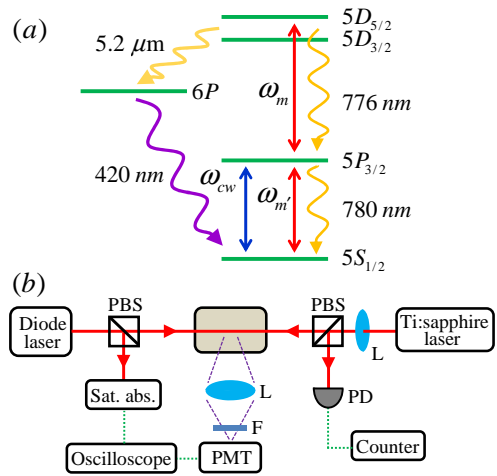}
  \caption {(a) (Color online) Schematic representation of the energy levels of Rb, where $\omega_{cw}$ represents the diode frequency and, $\omega_{m}$ and $\omega_{m\prime}$ are two distinct modes of the frequency comb. (b) Experimental setup. The components are the following: PBS, polarizing beam splitter; PMT, photomultiplier tube; PD, photodetector; L, lens; F, filter.}
  \end{figure} 

A simplified scheme of the experimental setup together with the relevant energy levels is presented in Fig.~1. 
A diode laser, stabilized in temperature and with a linewidth of about $1$~MHz, is used to excite 
the $5S_{1/2} \longrightarrow 5P_{3/2}$ transition at $780$~nm. A train of fs pulses generated by a mode-locked Ti:sapphire laser (BR Labs Ltda) can excite both $5S_{1/2} \longrightarrow 5P_{3/2}$ and $5P_{3/2} \longrightarrow 5D$ transitions. The two beams 
are overlapped, with orthogonal linear polarizations and in a counterpropagating configuration, 
in the center of a sealed Rb vapor cell. The vapor cell is heated to 
$\approx 80~^0$C and contains both $^{85}$Rb and $^{87}$Rb isotopes in their natural abundances. 

The Ti:sapphire laser produces $100$~fs pulses and $300$~mW of average power, such that the power
per mode is $\approx 60$~$\mu$W. The $f_{R}=1$~GHz repetition rate is measured with a photodiode and
phase locked to a signal generator (E8663B-Agilent), with $1$ Hz resolution, while the
carrier-envelope-offset frequency, $f_{0}$, is left free. The diode laser can sweep over
$10$~GHz by tuning its injection current and a saturated absorption setup is used to
calibrate its frequency. A direct detection of the diode-beam transmission after passing through
the cell gives information about the absorption in the $5S \longrightarrow 5P$ transition. The diameter of the
two beams at the center of the cell is on the order of $250$~$\mu$m for the fs beam and $1.8$~mm for
the diode laser, which lead to a common interaction time of about $800$~ns (corresponding to a linewidth of $\gamma\approx 2\pi\times200$~kHz).

\vspace{0.0cm}
\begin{figure}[htbp]
  \centering
  \includegraphics[width=7cm]{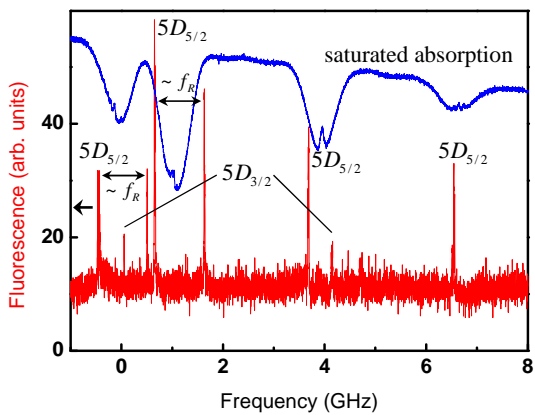}
  \vspace{0.0cm}  
  \caption{(Color online) Fluorescence from the $6P_{3/2} \longrightarrow 5S_{1/2}$ decay as a function of the diode laser frequency for the four $D_{2}$ Doppler lines. The saturated absorption signal (upper curve) is detected simultaneously.}
    \end{figure}
\vspace{0.0cm}

The fluorescence at 420 nm emitted by spontaneous decay from 6P state to 5S is collected at $90^0$, 
using a $10$~cm focal lens and a filter to cut the scattered light from the excitation laser beams.
The signal is detected with a photomultiplier tube and recorded on a digital oscilloscope.
Fig. 2 shows the fluorescence signal (lower curve), for a fixed $f_{R}$, as the 
diode frequency is scanned over the four Doppler-broadened $D_{2}$ lines of the $^{85}$Rb and 
$^{87}$Rb. The spectrum consists of several narrow peaks over a flat background. The 
narrow peaks are due to the two-photon transition excited by both lasers: the diode laser and the 
different modes of the frequency comb; while the background is due only to excitation by
the frequency comb. In the same scan we can observe, simultaneously, peaks associated to the 
excitation of the $5D_{3/2}$ and $5D_{5/2}$ states. We also see two peaks, separated by one 
$f_{R}$ in optical frequency of the diode laser, that correspond to the same 
transition excited by two neighboring modes of the frequency comb. 
The saturated absorption curve (upper curve) is used to calibrate the diode frequency.

\vspace{0.0cm}
\begin{figure}[htbp]
  \centering
  \includegraphics[width=8.0cm]{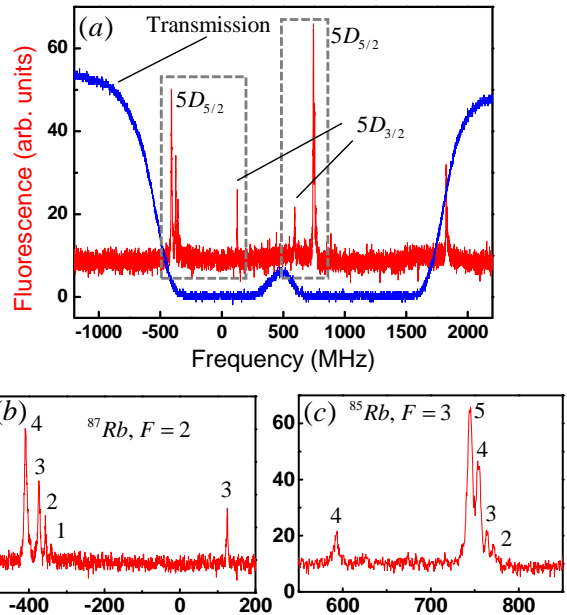}
  \vspace{0.0cm}  
  \caption{(Color online) (a) Fluorescence at $420$~nm as a function of the diode laser frequency. The diode transmission after the Rb cell (blue line) is detected simultaneously. Zooms of the regions inside the two rectangles in (a) for excitation from the hyperfine ground states (b) $F=2$ of $^{87}$Rb and (c) $F=3$ of $^{85}$Rb. The numbers at the peaks denote the hyperfine final state of the transition.}
    \end{figure}

The fluorescence for the Doppler lines $F=2$ of $^{87}$Rb and $F=3$ of $^{85}$Rb, as a function of the diode frequency, 
is shown in more detail in Fig. 3. We also present a measurement of the
direct transmission of the diode beam after passing through the warm cell (blue curve), 
indicating the strong absorption in the center of the Doppler lines. 
Figures 3(b) and 3(c) display a zoom of the two frequency regions selected in Fig. 3(a). As we can see, 
each peak in Fig. 2 consists of a group of peaks, which are the result of excitation from a given ground-state 
hyperfine level to all the hyperfine levels ($F^{\prime\prime}$) of the excited-state.

We can understand and calculate such spectra using a simple model consisting of independent three-level 
cascade systems interacting with two cw lasers. As we are interested in the combined action of the 
diode and the fs lasers, we will neglect the background. We denote one of the hyperfine levels 
of the ground ($5S_{1/2},F$), intermediate ($5P_{3/2},F^{\prime}$), and 
final ($5D,F^{\prime\prime}$) states as $\left|1\right\rangle$, $\left|2\right\rangle$, and $\left|3\right\rangle$, respectively.
The train of ultrashort pulses is described in the frequency domain as a frequency comb and 
we take into account only the modes that are near resonance with the 
$\left|2\right\rangle$$\rightarrow$$\left|3\right\rangle$ transitions. We also consider that the transitions
$\left|1\right\rangle$$\rightarrow$$\left|2\right\rangle$ are only driven by the diode laser. As the diode beam may be intense, we cannot apply second-order 
time-dependent perturbation theory, as in Ref.~\cite{Stalnaker2010}. The calculations are performed from the full Bloch equations for a three-level 
cascade system. The $\left|1\right\rangle$$\rightarrow$$\left|2\right\rangle$ 
transitions may be open or closed, depending on which $F^{\prime}$ is being excited. However, due to the longer lifetimes of the $5D$ states (resulting in weaker optical pumping) and to simplify the calculations, we consider that the 
$\left|2\right\rangle$$\rightarrow$$\left|3\right\rangle$ transitions are always closed. 

For a specific transition $F$$\rightarrow$$F^{\prime}$$\rightarrow$$F^{\prime\prime}$, we calculate 
the population $\rho_{33}$ of the final $F^{\prime\prime}$ state in the steady-state regime by solving the Bloch
equation for one specific diode frequency ($\delta$) integrated over 
the contribution of all velocity groups ($\Delta$) within the Doppler profile:
\begin{equation}
\begin{split}
\rho^{(F,F',F'')}_{33}(&\delta) = \frac{1}{\left( 0.36\pi\Delta^2_D \right)^{1/2}}\\
\times \int^{\infty}_{-\infty} &\rho_{33}(\Delta,\delta) e^{- \Delta^2/0.36\Delta^2_D} d \Delta,\\
\end{split}
\end{equation}
with $\Delta_D$ being the inhomogeneous Doppler linewidth. In the calculations, the strength of a specific $F_i \rightarrow F_j$  transition is parametrized by the corresponding Rabi frequency: $\Omega_{F_i,F_j} = ({\mu_{F_i,F_j} E})/{\hbar}$,
where $\mu_{F_i,F_j}$ is the electric-dipole moment for the transition and in the direction 
of the electric field, whose amplitude $E$ may refer to the diode laser or a single
mode of the frequency comb, depending on the transition. As we do not know the carrier-envelope-offset frequency, 
the frequency of the mode that drives the $F^{\prime}$$\rightarrow$$F^{\prime\prime}$ transition 
is determine by the two-photon resonance condition combined with the diode frequency and the velocity of the atoms
at resonance. 

\vspace{0.0cm}
\begin{figure}[htbp]
  \centering
  \includegraphics[width=8.0cm]{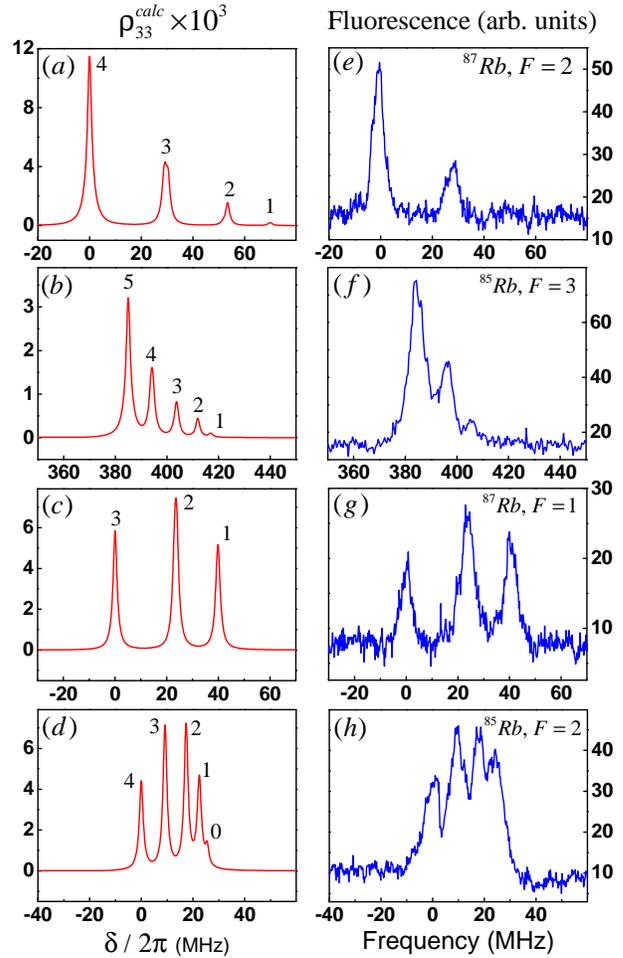}
  \vspace{0.0cm}  
  \caption{(a)-(d) Calculated population of level $\left|3\right\rangle$, $\rho^{calc}_{33}(\delta)$, and (e)-(h) Experimental 
	results for the fluorescence at $420$~nm, as a function of the diode frequency, for the four $D_{2}$ Doppler lines.}
    \end{figure}

To compare with the experimental spectra we need to add the contributions 
due to all allowed two-photon transitions: $\rho^{calc}_{33}(\delta) = \sum_{F,F',F''} \rho^{(F,F',F'')}_{33}(\delta)$. 
We consider that the two beams have orthogonal polarizations and sum all contributions weighted by the corresponding eletric-dipole moments. We also take into account that more than one velocity group contribute to the fluorescence from each two-photon transition~\cite{Liao1976,Stalnaker2010}. Figure 4 shows the calculated population $\rho^{calc}_{33}(\delta)$ of level $\left|3\right\rangle$ for four different groups of peaks at each of the $D_2$ Doppler lines. At the right column of Fig. 4 we display the experimental results obtained for the same 
conditions used in the calculations. As we can see, a good description of the experimental results is obtained, although the peaks appear wider than the calculated ones. We attribute the larger widths of the experimental peaks mainly to drifts, during the time of averaging of the experimental curves, of the offset frequency $f_0$ of the frequency comb, and to the linewidth of the diode laser itself.

In conclusion, we show that the combined action of a cw laser and a train of ultra-short pulses can be used to drive two-photon transitions efficiently and with high spectral resolution. Although the experimental resolution is limited by the free running offset frequency and the linewidth of the diode laser, we are able to investigate simultaneously a large spectral range including the $5D_{3/2}$ and $5D_{5/2}$ states with good spectral resolution, and without the need of direct spectral filtering of the fs laser. A direct, quantitative description of the experimental spectra is given in the frequency domain picture, highlighting all physical elements that are important to explain our results.

This work was supported by CNPq, FACEPE and CAPES (Brazilian Agencies).

\end{document}